

%

%
 \font\twelvebf=cmbx12
 \font\twelvett=cmtt12
 \font\twelveit=cmti12
 \font\twelvesl=cmsl12
 \font\twelverm=cmr12		\font\ninerm=cmr9
 \font\twelvei=cmmi12		\font\ninei=cmmi9
 \font\twelvesy=cmsy10 at 12pt	\font\ninesy=cmsy9
 \skewchar\twelvei='177		\skewchar\ninei='177
 \skewchar\seveni='177	 	\skewchar\fivei='177
 \skewchar\twelvesy='60		\skewchar\ninesy='60
 \skewchar\sevensy='60		\skewchar\fivesy='60
%
%

%
 \font\fourteenrm=cmr12 scaled 1200
 \font\seventeenrm=cmr12 scaled 1440
 \font\fourteenbf=cmbx12 scaled 1200
 \font\seventeenbf=cmbx12 scaled 1440
%
%

%
%
%
\font\tenmsb=msbm10
\font\twelvemsb=msbm10 scaled 1200
\newfam\msbfam

%
\font\tensc=cmcsc10
\font\twelvesc=cmcsc10 scaled 1200
\newfam\scfam

%
\def\seventeenpt{\def\rm{\fam0\seventeenrm}%
 \textfont\bffam=\seventeenbf	\def\bf{\fam\bffam\seventeenbf}}
\def\fourteenpt{\def\rm{\fam0\fourteenrm}%
 \textfont\bffam=\fourteenbf	\def\bf{\fam\bffam\fourteenbf}}
\def\twelvept{\def\rm{\fam0\twelverm}%
 \textfont0=\twelverm	\scriptfont0=\ninerm	\scriptscriptfont0=\sevenrm
 \textfont1=\twelvei	\scriptfont1=\ninei	\scriptscriptfont1=\seveni
 \textfont2=\twelvesy	\scriptfont2=\ninesy	\scriptscriptfont2=\sevensy
 \textfont3=\tenex	\scriptfont3=\tenex	\scriptscriptfont3=\tenex
 \textfont\itfam=\twelveit	\def\it{\fam\itfam\twelveit}%
 \textfont\slfam=\twelvesl	\def\sl{\fam\slfam\twelvesl}%
 \textfont\ttfam=\twelvett	\def\tt{\fam\ttfam\twelvett}%
 \scriptfont\bffam=\tenbf 	\scriptscriptfont\bffam=\sevenbf
 \textfont\bffam=\twelvebf	\def\bf{\fam\bffam\twelvebf}%
 \textfont\scfam=\twelvesc	\def\sc{\fam\scfam\twelvesc}%
 \textfont\msbfam=\twelvemsb	
 \baselineskip 14pt%
 \abovedisplayskip 7pt plus 3pt minus 1pt%
 \belowdisplayskip 7pt plus 3pt minus 1pt%
 \abovedisplayshortskip 0pt plus 3pt%
 \belowdisplayshortskip 4pt plus 3pt minus 1pt%
 \parskip 3pt plus 1.5pt
 \setbox\strutbox=\hbox{\vrule height 10pt depth 4pt width 0pt}}
\def\tenpt{\def\rm{\fam0\tenrm}%
 \textfont0=\tenrm	\scriptfont0=\sevenrm	\scriptscriptfont0=\fiverm
 \textfont1=\teni	\scriptfont1=\seveni	\scriptscriptfont1=\fivei
 \textfont2=\tensy	\scriptfont2=\sevensy	\scriptscriptfont2=\fivesy
 \textfont3=\tenex	\scriptfont3=\tenex	\scriptscriptfont3=\tenex
 \textfont\itfam=\tenit		\def\it{\fam\itfam\tenit}%
 \textfont\slfam=\tensl		\def\sl{\fam\slfam\tensl}%
 \textfont\ttfam=\tentt		\def\tt{\fam\ttfam\tentt}%
 \scriptfont\bffam=\sevenbf 	\scriptscriptfont\bffam=\fivebf
 \textfont\bffam=\tenbf		\def\bf{\fam\bffam\tenbf}%
 \textfont\scfam=\tensc		\def\sc{\fam\scfam\tensc}%
 \textfont\msbfam=\tenmsb	
 \baselineskip 12pt%
 \abovedisplayskip 6pt plus 3pt minus 1pt%
 \belowdisplayskip 6pt plus 3pt minus 1pt%
 \abovedisplayshortskip 0pt plus 3pt%
 \belowdisplayshortskip 4pt plus 3pt minus 1pt%
 \parskip 2pt plus 1pt
 \setbox\strutbox=\hbox{\vrule height 8.5pt depth 3.5pt width 0pt}}

%
\def\twelvepoint{%
 \def\small{\tenpt\rm}%
 \def\normal{\twelvept\rm}%
 \def\big{\fourteenpt\rm}%
 \def\huge{\seventeenpt\rm}%
 \footline{\hss\twelverm\folio\hss}
 \normal}
%

%
\def\bigbold{\big\bf}

%
\catcode`\@=11
%
%
\def\footnote#1{\edef\@sf{\spacefactor\the\spacefactor}#1\@sf
 \insert\footins\bgroup\small
 \interlinepenalty100	\let\par=\endgraf
 \leftskip=0pt		\rightskip=0pt
 \splittopskip=10pt plus 1pt minus 1pt	\floatingpenalty=20000
 \smallskip\item{#1}\bgroup\strut\aftergroup\@foot\let\next}
%
%
%
%
\def\hexnumber@#1{\ifcase#1 0\or 1\or 2\or 3\or 4\or 5\or 6\or 7\or 8\or
 9\or A\or B\or C\or D\or E\or F\fi}
\edef\msbfam@{\hexnumber@\msbfam}

%
%
%
\catcode`\@=12

\newcount\EQNO      \EQNO=0
\newcount\FIGNO     \FIGNO=0
\newcount\REFNO     \REFNO=0
\newcount\SECNO     \SECNO=0
\newcount\SUBSECNO  \SUBSECNO=0
\newcount\FOOTNO    \FOOTNO=0
\newbox\FIGBOX      \setbox\FIGBOX=\vbox{}
\newbox\REFBOX      \setbox\REFBOX=\vbox{}
\newbox\RefBoxOne   \setbox\RefBoxOne=\vbox{}

\expandafter\ifx\csname normal\endcsname\relax\def\normal{\null}\fi

\def\Eqno{\global\advance\EQNO by 1 \eqno(\the\EQNO)%
    \gdef\label##1{\xdef##1{\nobreak(\the\EQNO)}}}
\def\Fig#1{\global\advance\FIGNO by 1 Figure~\the\FIGNO%
    \global\setbox\FIGBOX=\vbox{\unvcopy\FIGBOX
      \narrower\smallskip\item{\bf Figure \the\FIGNO~~}#1}}
\def\Ref#1{\global\advance\REFNO by 1 \nobreak[\the\REFNO]%
    \global\setbox\REFBOX=\vbox{\unvcopy\REFBOX\normal
      \smallskip\item{\the\REFNO .~}#1}%
    \gdef\label##1{\xdef##1{\nobreak[\the\REFNO]}}}
\def\Section#1{\SUBSECNO=0\advance\SECNO by 1
    \bigskip\leftline{\bf \the\SECNO .\ #1}\nobreak}
\def\Subsection#1{\advance\SUBSECNO by 1
    \medskip\leftline{\bf \ifcase\SUBSECNO\or
    a\or b\or c\or d\or e\or f\or g\or h\or i\or j\or k\or l\or m\or n\fi
    )\ #1}\nobreak}
\def\Footnote#1{\global\advance\FOOTNO by 1 
    \footnote{\nobreak$\>\!{}^{\the\FOOTNO}\>\!$}{#1}}
\def\SameFootnote{$\>\!{}^{\the\FOOTNO}\>\!$}

\def\References{\bigskip\centerline{\bf REFERENCES}
                \smallskip\copy\REFBOX}
\def\NewRefPage{\setbox\RefBoxOne=\vbox{\unvcopy\REFBOX}
		\setbox\REFBOX=\vbox{}
		\def\References{\bigskip\centerline{\bf REFERENCES}
                		\nobreak\smallskip\nobreak\copy\RefBoxOne
				\vfill\eject
				\smallskip\copy\REFBOX}
		\def\NewRefPage{}}




\font\twelvebm=cmmib10 at 12pt
\font\tenbm=cmmib10
\font\ninei=cmmi9
\newfam\bmfam

\def\twelvepointbmit{
\textfont\bmfam=\twelvebm
\scriptfont\bmfam=\ninei
\scriptscriptfont\bmfam=\seveni
\def\bmit{\fam\bmfam\twelvebm}
}

\def\tenpointbmit{
\textfont\bmfam=\tenbm
\scriptfont\bmfam=\seveni
\scriptscriptfont\bmfam=\fivei
\def\bmit{\fam\bmfam\tenbm}
}

\tenpointbmit

\mathchardef\Gamma="7100
\mathchardef\Delta="7101
\mathchardef\Theta="7102
\mathchardef\Lambda="7103
\mathchardef\Xi="7104
\mathchardef\Pi="7105
\mathchardef\Sigma="7106
\mathchardef\Upsilon="7107
\mathchardef\Phi="7108
\mathchardef\Psi="7109
\mathchardef\Omega="710A
\mathchardef\alpha="710B
\mathchardef\beta="710C
\mathchardef\gamma="710D
\mathchardef\delta="710E
\mathchardef\epsilon="710F
\mathchardef\zeta="7110
\mathchardef\eta="7111
\mathchardef\theta="7112
\mathchardef\iota="7113
\mathchardef\kappa="7114
\mathchardef\lambda="7115
\mathchardef\mu="7116
\mathchardef\nu="7117
\mathchardef\xi="7118
\mathchardef\pi="7119
\mathchardef\rho="711A
\mathchardef\sigma="711B
\mathchardef\tau="711C
\mathchardef\upsilon="711D
\mathchardef\phi="711E
\mathchardef\cho="711F
\mathchardef\psi="7120
\mathchardef\omega="7121
\mathchardef\varepsilon="7122
\mathchardef\vartheta="7123
\mathchardef\varpi="7124
\mathchardef\varrho="7125
\mathchardef\varsigma="7126
\mathchardef\varphi="7127



%
%
\twelvepoint			
%
%



\font\twelvebm=cmmib10 at 12pt
\font\tenbm=cmmib10
\font\ninei=cmmi9
\newfam\bmfam

\def\twelvepointbmit{
\textfont\bmfam=\twelvebm
\scriptfont\bmfam=\ninei
\scriptscriptfont\bmfam=\seveni
\def\bmit{\fam\bmfam\twelvebm}
}

\def\tenpointbmit{
\textfont\bmfam=\tenbm
\scriptfont\bmfam=\seveni
\scriptscriptfont\bmfam=\fivei
\def\bmit{\fam\bmfam\tenbm}
}

\tenpointbmit

\mathchardef\Gamma="7100
\mathchardef\Delta="7101
\mathchardef\Theta="7102
\mathchardef\Lambda="7103
\mathchardef\Xi="7104
\mathchardef\Pi="7105
\mathchardef\Sigma="7106
\mathchardef\Upsilon="7107
\mathchardef\Phi="7108
\mathchardef\Psi="7109
\mathchardef\Omega="710A
\mathchardef\alpha="710B
\mathchardef\beta="710C
\mathchardef\gamma="710D
\mathchardef\delta="710E
\mathchardef\epsilon="710F
\mathchardef\zeta="7110
\mathchardef\eta="7111
\mathchardef\theta="7112
\mathchardef\iota="7113
\mathchardef\kappa="7114
\mathchardef\lambda="7115
\mathchardef\mu="7116
\mathchardef\nu="7117
\mathchardef\xi="7118
\mathchardef\pi="7119
\mathchardef\rho="711A
\mathchardef\sigma="711B
\mathchardef\tau="711C
\mathchardef\upsilon="711D
\mathchardef\phi="711E
\mathchardef\cho="711F
\mathchardef\psi="7120
\mathchardef\omega="7121
\mathchardef\varepsilon="7122
\mathchardef\vartheta="7123
\mathchardef\varpi="7124
\mathchardef\varrho="7125
\mathchardef\varsigma="7126
\mathchardef\varphi="7127



\vskip 2cm 

\centerline{\bigbold NON-RIEMANNIAN GRAVITY}\vskip 0.7cm
\centerline{\bigbold AND THE}\vskip 0.7cm
\centerline{\bigbold EINSTEIN-PROCA SYSTEM}\vskip 0.7cm
\bigskip\bigskip\bigskip

\centerline{T Dereli}
\medskip
\centerline{\it Department of Physics,}
\centerline{\it Middle East Technical University,
		Ankara, Turkey}
\centerline{\tt tdereli@rorqual.cc.metu.edu.tr}

\medskip
\medskip
\medskip

\centerline{M \"Onder}
\medskip
\centerline{\it Department of Physics,}
\centerline{\it Haceteppe University,
		Ankara, Turkey}

\centerline{\tt onder@eti.cc.hun.edu.tr}

\vskip 1cm

\centerline{J\"org Schray}
\medskip
\centerline{Robin W Tucker}
\medskip
\centerline{Charles Wang}
\medskip

\centerline{\it School of Physics and Chemistry,}
\centerline{\it University of Lancaster,
		Bailrigg, Lancs. LA1 4YB, UK}
\medskip
\centerline{\tt r.tucker{\rm @}lancaster.ac.uk}
\centerline{\tt j.schray{\rm @}lancaster.ac.uk}
\centerline{\tt c.wang{\rm @}lancaster.ac.uk}

\vskip 1cm

\vskip 2cm


\centerline{\bf ABSTRACT}
\vskip 1cm

\midinsert
\narrower\narrower\noindent


We argue that  {\it all}\ Einstein-Maxwell or Einstein-Proca solutions
to general relativity may  be used to construct a large class of solutions
(involving torsion and non-metricity) to theories of non-Riemannian 
gravitation that have been recently discussed in the literature.

\endinsert




\vfill
\eject

\headline={\hss\rm -~\folio~- \hss}     

\def\frac#1#2{{#1\over #2}}

\Section{Introduction}

Non-Riemannian geometries feature in a number of theoretical descriptions
of the interactions between fields and gravitation. Since the early pioneering
 work by Weyl, Cartan, Schroedinger and others such geometries have often
provided a succinct and elegant guide towards the search for unification of
the forces of nature
\Ref{ H Weyl, Geometrie und Physik, 
  Naturwissenschaften {\bf 19} (1931) 49}\label\weylref 
. In recent times interactions with supergravity have
been encoded into torsion fields induced by spinors and dilatonic
interactions from low energy effective string theories 
have been encoded into connections
that are not metric-compatible 
\Ref{J Scherk, J H Schwarz, Phys. Letts {\bf 52B} (1974) 347},
\Ref{ T Dereli, R W Tucker, An Einstein-Hilbert Action for Axi-Dilaton
Gravity in \break 4-Dimensions, Lett. Class. Q. Grav. {\bf 12} (1995) L31},
\Ref{ T Dereli, M \"Onder,  R W Tucker, Solutions for Neutral  Axi-Dilaton
Gravity in 4-Dimensions, Lett. Class. Q. Grav. {\bf 12} (1995) L25},
\Ref{ T Dereli, R W Tucker,   Class. Q. Grav. {\bf 11} (1994) 2575}.
However theories in which the
non-Riemannian geometrical fields are dynamical in the absence of matter
are more elusive to interpret. It has been suggested 
that they may play an important role in certain astrophysical contexts  
\Ref{R W Tucker, C Wang, 
Class. Quan. Grav. {\bf 12} (1995) 2587}\label\newgravity.
Part of the difficulty in interpreting such fields is that there is little
experimental guidance available for the construction of a viable
theory that can compete effectively with general relativity in domains that
are currently accessible to observation. In such circumstances one must be
guided by the classical solutions admitted by theoretical models that admit
dynamical non-Riemannian structures
\Ref{F W Hehl, J D McCrea, E W Mielke, Y Ne'eman: ``Metric-affine
gauge theory of gravity: field equations, Noether identities, 
world
spinors, and breaking of dilation invariance''.  
Physics Reports, {\bf 258} (1995) 1.
}\label\Hehl,
\Ref{ F W Hehl, E Lord, L L Smalley, Gen. Rel. Grav. {\bf 13} (1981) 1037},
\Ref{P Baekler,  F W Hehl, E W Mielke  ``Non-Metricity and Torsion'' in
Proc. of 4th Marcel Grossman Meeting on General Relativity, Part A,
Ed. R Ruffini  (North Holland 1986) 277},
\Ref{V N Ponomariev, Y Obukhov, Gen. Rel. Grav. {\bf 14} (1982) 309},
\Ref{ J D McCrea, Clas. Q. Grav. {\bf 9} (1992) 553}.
A number of recent papers have pursued this approach and have found static
spherically symmetric solutions to particular models
\Ref{R Tresguerres, 
Z. f\"ur Physik {\bf C} 65 (1995) 347 }\label\TresA,\ 
\Ref{R Tresguerres, 
Phys.Lett. {\bf A200} 65 (1995) 405}\label\TresB.
In 
\newgravity\  it was pointed
out that in a particularly simple model all Einstein-Maxwell solutions
to general relativity could be  used to generate dynamic non-Riemannian
geometries and a tentative interpretation was offered for the matter
couplings in such a model. Particular solutions have also been
found to more complex models, provided the coupling constants 
 in the action are correlated \TresA,\ \TresB,\    
\Ref{Y Obukhov, E J Vlachynsky, W Esser, R Tresguerres, F W Hehl. An exact
solution of the metric-affine gauge theory with dilation, shear and spin
charges, gr-qc 9604027 (1996)}\label\TresC,\  
\Ref{ E J Vlachynsky, R Tresguerres, Y Obukhov,  F W Hehl. 
An axially symmetric
solution of metric-affine gravity, 
gr-qc 9604035 (1996)}\label\TresD.\ 
  It is the purpose of
this note to point out that, 
 if such correlations are maintained
then  solutions may be generated from {\it all Einstein-Maxwell solutions of
general relativity}. Furthermore 
 the correlations may be  discarded and    solutions  generated from 
{\it all Einstein-Proca solutions of
general relativity} with or without the inclusion of a cosmological term in
the action.



\def\und{\underbrace}

\twelvepointbmit

\def\a{\alpha\,}
\def\b{\beta\,}

\def\g{{\bmit\gamma}}

\def\wd{\wedge}

\def\R#1#2{R^#1{}_#2}

\def\om#1#2{\omega^#1{}_#2}

\def\e#1{e^#1{}}

\def\frac#1#2{{#1\over #2}\,}

\def\k{\kappa\,}

\def\dott{\dot{\,}}
\def\dQ{d\,Q}
\def\var#1#2{ \und{(#1) }_{#2}\dott}

\def\RR{{\cal R}}

\Section{Non-Riemannian Geometry}

To establish notation (which follows
\Ref{ I M Benn, R W Tucker, {\bf An Introduction to Spinors and Geometry
with
Applications in Physics}, (Adam Hilger) (1987)}\label\book) 
we briefly recall some basic definitions.
A non-Riemannian geometry is specified by a metric tensor field ${\bf g}$ and a
linear connection $\nabla$ on the bundle of linear frames over a manifold.
In a local coframe $\{\e a\}$ with dual frame $\{X_b\}$ such that 
$\e a(X_b)=\delta^a{}_b$, the connection forms $\om ab$ satisfy
$ \om cb (X_a)\equiv \e c(\nabla_{X_a} X_b)$.
 The connection is not metric
compatible when the tensor field ${\bf S}=\nabla {\bf g}$
is non zero. In the following we use local orthonormal  frames
so ${\bf g}=\eta_{ab}\, \e a\otimes \e b$, 
($\eta_{ab}=$diag$(-1,1,1,1,\ldots)$),
non-metricity 1-forms 
$Q_{ab}\equiv {\bf S}(-,X_a,X_b)$,  torsion 2-forms
$T^a\equiv d\,\e a +\om ab\wd \e b$
 and  curvature 2-forms
$\R ab \equiv  d\,\om ab + \om ac \wd \om cb $. The general curvature scalar
$\RR$  associated with the connection 
is defined by
 $\RR *1\equiv \R ab\wd *(e_a\wd e^b)$
in terms of the Hodge operator for the metric.

\Section{Field Equations for Non-Riemannian Gravity}

We begin by analysing the theory derived from the action functional 
$$
\Lambda[e,\omega]=\k \RR *1 +\frac\a2 d\,Q\wd * d\,Q
+\frac\b2 Q\wd * Q + \frac\g2 T\wd * T \Eqno 
$$\label\LLambda
where
$\kappa, \a,\b,\g$ are real couplings, $Q =\eta^{ab}\,Q_{ab}$
and $T=i_c T^c$. (Here and below $i_c\equiv i_{X_c}$ in terms if the
contraction operator.)
For any form ${\cal T}$ we denote its variation induced by a coframe 
  variation $\{\dot{e}^a\}$
 and a connection variation $\{\dot{\omega}^a{}_b\}$  by 
$\und{{\cal T}}_e\dot{\,} $ and
$\und{{\cal T}}_\omega\dot{\,} $ respectively.
Consider orthonormal coframe  induced variations of $\Lambda[e,\omega]$. 
The first term gives
$$
\und{ (\R ab\wd *(e_a\wd e^b))}_e\dott =\dot\e c \wd \R ab\wd *(e_a\wd \e
b\wd e_c). \Eqno
$$
Since $\var \dQ e =0$  and  $\var Q e =0$ it follows that
$$
\var {\frac12\dQ \wd \dQ}{e}= \frac12\dot{\e a}  
\wd (\dQ\wd i_a * \dQ -i_a\dQ * \dQ) \Eqno
$$
$$
\var {\frac12 Q \wd Q}{e}= 
-\frac12\dot{\e a}  \wd (Q\wd i_a * Q +i_a Q * Q) .\Eqno
$$
The coframe variation of the last term may be computed by noting that
$\var {T^a} e = D\,\dot\e a$ and $i_b\dot\e a=-i_{\dot{X_b}}\e a$.
Thus
$\var T e =\var {i_{X_a} T^a} e =i_{\dot{X_a}}\, T^a +i_a\var{T^a}{e}$
and
$$
\frac12\var{ T\wd * T}{e}= \dot{\e a}\wd\{  i_k(T^k \wd * (T\wd \e
a))-D\,(i_a *T) -\frac12  (T\wd i_a *T + i_aT \wd *T)\}. \Eqno
$$
Turning to the connection induced variations of $\Lambda[e,\omega]$ 
it follows from the definition of the curvature forms that
$$
\var{\RR *1}{\omega}=
\dot{\omega}^a{}_b\,\wd D\,*(e_a\wd \e b).
$$
Next we note that 
$
\var Q
\omega=\eta^{ab}\var{Q_{ab}}{\omega}=
g^{ab}\var{D\,\eta_{ab}}{\omega}=-2\dot{\omega}
^a{}_b\,\delta^b{}_a
$
hence
$$
\frac12\var {Q\wd * Q}{\omega}=
\dot{Q}\wd  *  Q
=-2\dot{\omega}^a{}_b \,\delta^a{}_b\wd  *  Q
\Eqno
$$
and
$$
\frac12\var {dQ\wd * \dQ}{\omega}=
\dot{Q}\wd d\, * d\, Q
=-2\dot{\omega}^a{}_b \,\delta^a{}_b\wd d\, * d\, Q
\quad  {\hbox{mod} } d.\Eqno
$$
Finally
$$\frac12\var{T\wd * \,T}{\omega}
=\frac12\var{i_a T^a \wd *\,i_b T^b}{\omega} 
=i_a \var{T^a}{\omega}\wd *\, i_b T^b \Eqno $$
but $\var{T^a}{\omega}= \dot{\omega}^a{}_b\,\wd \e b$
so
$$\frac12\var{T\wd * \,T}{\omega}
=i_a(\dot{\omega}^a{}_c\,\wd \e c)\wd *i_b\,T^b=
-\dot{\omega}^a{}_b\,\wd \e b \wd i_a \,* T. \Eqno
$$
Thus the variational condition $\var{\Lambda}{\omega}=0 \quad
 {\hbox{mod }} d$ yields the field equation:
$$
\k D\,*(e_a\wd \e b)=2\delta^b{}_a\,(\a d\,*d\,Q+\b*Q) + \g \e b\wd i_a *\,T
.\Eqno
$$
It is instructive to take the trace of this equation and replace it by the
equivalent equations:
$$ \a d\,*d\, Q+\b *Q=\frac{\g(1-n)}{2n}*T \Eqno$$\label\LambdaCartaneqnsone
$$ 
\k D\,*(e_a\wd \e b)=\delta^b{}_a  \frac{(1-n)}{n}\g *T +\g \e b \wd i_a
\,*T\Eqno
$$\label\LambdaCartaneqnstwo
where $n$ is the dimension of the manifold.

The variational condition $\var{\Lambda}{e}=0 \quad {\hbox{mod }} d$
yields the Einstein field equation:
$$
\k\R ab\wd *(e_a\wd \e b \wd e_c)
+\tau_c[\a]
+\tau_c[\b]
+\tau_c[\g]=0
\Eqno
$$\label\LambdaEineqns
where
$$
\tau_c[\a]=\frac{\a}{2}(d\,Q \wd i_c *d\, Q-i_c\,d\, Q\wd * d\,Q)
\Eqno
$$
$$
\tau_c[\b]
=-\frac{\b}{2} (Q \wd i_c * Q+i_c\, Q\wd * Q)
\Eqno
$$
$$
\tau_c[\g]
=\g\{i_k(T^k \wd *(T\wd e_c))-Di_c*T -\frac12(T\wd i_c *T+i_cT\wd *T)\}
.\Eqno
$$

\Section{Non-Riemannian Solutions Generated from  the Einstein-Maxwell System}
\def\AA#1{{\cal  A}_{#1}}

We look for  solutions to these field equations in which the
torsion and non-metricity forms are expressed in terms of general 1-forms
$\AA1,\AA2$ and $\AA3$ on spacetime,  according to the ansatz:
$$
Q_{ab}= e_a\, i_b \AA1 + e_b\, i_a \AA1 -\frac2n \eta_{ab} \AA1  
-\frac{\eta_{ab}}{n} \AA2 \Eqno
$$\label\Qansatz
$$
T^a=\frac{1}{2n} \e a \wd \AA3
.\Eqno 
$$\label\Tansatz
In the following we are concerned with spacetime solutions and set $n=4$.
This ansatz ensures that a remarkable cancellation occurs in the field
equations
and is responsible for the particular solutions that have been discussed in
\TresC,\ \TresD. To describe the nature of the solutions that arise 
we introduce an action
functional of vacuum Einstein-Maxwell type:
$$
\Lambda_{E-M}[{\bf g},A]=\k \RR*1+\frac{\a}{2} d\,A \wd * d\, A \Eqno
$$
where $F=d\,A$ is the Maxwell 2-form.
The traditional field equations of Einstein-Maxwell type
 follow from \LambdaEineqns\  by putting $\b=\g=0$ and setting the
forms $\om ab$ to define the metric compatible torsion free
 Levi-Civita connection.
(The coupling $\a $  is negative for the physical Einstein-Maxwell system.)
As a result of a somewhat tedeious calculation (which has been checked
using the programme Manifolds
\Ref{R W Tucker, C Wang, {\it Manifolds: A Maple Package for Differential
Geometry} (1996)}
 in the symbolic algebra language Maple)
the following result emerges.

Given any solution $({\bf g}={\bf g}_{E-M}, A=A_{E-M})$ 
of  Einstein-Maxwell  type to the
field equations generated from the traditional  action 
$\Lambda_{E-M}[{\bf g},A]$ above, then a solution to the field equations
\LambdaEineqns,\ \LambdaCartaneqnsone,\  \LambdaCartaneqnstwo\  
generated
by the action $\Lambda[e,\omega]$  is given by 
${\bf g}= {\bf g}_{E-M}$
together with
the  
ansatz \Qansatz,\ \Tansatz\   
 where the 1-forms $\AA1 , \AA2 , \AA3$ are given by
$$ \AA1=-\frac{2\b}{3\k} A_{E-M} \Eqno$$
$$ \AA2= A_{E-M} \Eqno$$
$$ \AA3=-\frac{64\b}{9\g} A_{E-M} \Eqno$$
provided the couplings in the action satisfy
$$ 12\b\g -9\g\k -64\b\k=0.\Eqno$$\label\couplingcondition 
Thus a result of \TresC\ arises as a particular solution generated 
from the class of {\it all vacuum Einstein-Maxwell solutions}. Furthermore
we next show  that  the constraint on the couplings in the action can be
removed, if necessary, by generating solutions from the class of all
{\it all vacuum Einstein-Proca solutions}.

\Section{Extended Action and Field Equations}

In this section we also broaden the theory by considering the action
functional: 
$$
\Lambda_1[e,\omega]=\Lambda[e,\omega]+\frac12\,\xi\, T_c\wd *\, T^c
\Eqno
$$ \label\extendedaction
where $\xi$ is an arbitrary real coupling.
The variations of the extra term follow from
$$
\frac12\var{T_c\wd *\, T^c}{e}=\var{T^c}{e}\wd *\, T_c
+\frac12\dot{\e a}\{T_c\wd i_a\,*\,T^c-i_a\,T_c\wd *\, T^c\}
$$
or
$$
\frac12\var{T_c\wd *\, T^c}{e}=\dot{\e c}\wd (D\,*T_c
+\frac12\{T_c\wd i_a\,*\,T^c-i_a\,T_c\wd *\, T^c\})
\Eqno
$$
and
$$
\frac12\var{T_c\wd *\, T^c}{\omega}=
\var{T^c}{\omega}\wd *\,T_c=\dot{\om cb}\wd \e b \wd *\, T_c.\Eqno
$$
Thus 
$\var{\Lambda_1}{e}=0 \quad \hbox{mod } d$
yields
$$
\k\R ab\wd *(e_a\wd \e b \wd e_c)
+\tau_c[\a]
+\tau_c[\b]
+\tau_c[\g]
+\tau_c[\xi]=0
\Eqno
$$\label\LambdaoneEineqns
where
$$
\tau_c[\xi]
= \xi(D\,*T_c
+\frac12\{T_c\wd i_a\,*\,T^c-i_a\,T_c\wd *\, T^c\})
\Eqno
$$
while $\var{\Lambda_1}{\omega}=0 \quad \hbox{mod } d$
gives
$$
\k D\,*(e_a\wd \e b)=2\delta^b{}_a\,(\a d\,*d\,Q+\b*Q) + \g \e b\wd i_a *\,T
-\xi \e b \wd *\, T_a
\Eqno
$$
or eqivalently:
$$ \a d\,*d\, Q+\b *Q=
\frac{\g(1-n)}{2n}*T +\frac{\xi}{2n} \e c\wd *\, T_c\Eqno
$$\label\LambdaoneCartaneqnsone
$$ 
\k D\,*(e_a\wd \e b)=
\delta^b{}_a \frac{(1-n)}{n}\g *T + \delta^b{}_a \frac{\xi}{n} \e c\wd *\, T_c
 +\g \e b \wd i_a \,*T  -\xi \e b \wd *\, T_a.\Eqno
$$\label\LambdaoneCartaneqnstwo

\Section{Non-Riemannian Solutions Generated from  the Einstein-Proca System}

To describe the nature of the solutions that arise to these equations with
the ansatz above we introduce an action functional
of vacuum Einstein-Proca type:
$$
\Lambda_{E-P}[{\bf g},A]=\k \RR*1+\frac{\a}{2} d\,A \wd * d\, A 
+\frac{\b-\beta_0}{2} A\wd *\, A
\Eqno
$$
where $A$ is the Proca 1-form.
The field equations of Einstein-Proca type
 follow from \LambdaEineqns\ 
by replacing $\b$ by $\b-\beta_0$ and setting $\g=0$ and 
 the
forms $\om ab$ to the metric compatible torsion free Levi-Civita ones.
(The constant ${\b-\beta_0}$  
is positive for the physical Einstein-Proca system
describing a massive vector field coupled to Einsteinian gravity.)
Given any solution $({\bf g}={\bf g}_{E-P}, A=A_{E-P})$ 
of Einstein-Proca  type to the
field equations generated from the traditional  action functional 
$\Lambda_{E-P}[{\bf g},A]$ above, then a solution to the field equations
\LambdaoneEineqns,\ \LambdaoneCartaneqnsone,\ \LambdaoneCartaneqnstwo\ 
 generated
by the action functional $\Lambda_1[e,\omega]$  is given by   
${\bf g}= {\bf g}_{E-M}$
together with
the  ansatz \Qansatz,\ \Tansatz\  where the 1 -forms 
$\AA1 , \AA2 , \AA3$ are given by
$$ \AA1=-\frac{2\b}{3\k} A_{E-P} \Eqno$$
$$ \AA2= A_{E-P} \Eqno$$
$$ \AA3=-\frac{64\b}{(9\g+3\xi)} A_{E-P} \Eqno$$
provided the parameter $\beta_0$ determining the mass term in the
 Einstein-Proca solution is choosen to be
$$\beta_0=\frac{3\k(\xi+3\g)}{4(\xi+3\g-16\k)}.\Eqno$$

Thus by choosing any Einstein-Proca solution associated with the appropriate
``mass parameter'' one can generate a solution to the above theory without
constraining any of the coupling constants $\k,\a,\b,\g,\xi$.


\def\qq{{\cal Q}}
\Section{Generalisations}

 We have briefly considered  further generalisations.
Firstly if the ansatz \Tansatz\  is replaced by
$$
T^a=\frac{1}{2n} \e a \wd \AA3 + \frac{1}{2n}*( \e a \wd \AA4)
\Eqno 
$$\label\starTansatz 
where $\AA4$ is a 1-form on spacetime, then
the above result provides a class of solutions with arbitrary $\AA4$ but
constrained couplings.
Thus
if any Einstein-Proca solution $({\bf g}_{E-P}, A_{E-P})$ is used to define
the 1-forms $\AA1, \AA2, \AA3 $ and ${\bf g}$ as above then a solution to the
theory generated by $\Lambda_1[e,\omega]$ can be generated provided the
couplings in the action satisfy the constraint:
$$\xi=\k\Eqno$$\label\couplingconditionA\ 
in addition to  \couplingcondition.
In such a solution the 1-form $\AA4$ remains arbitrary. 
We note further that one may also include a cosmological constant in all the
actions above and generate solutions from either Einstein-Maxwell or
Einstein-Proca solutions in the presence of a cosmological constant.

While this work was in progress we became aware of the papers
\TresC\ and \TresD\  
in which particular solutions to a class of more general actions are
presented.
The authors study the theory based on the  action functional
$$
\Lambda_2=
{1\over2\kappa}\left[
a_0\,R^{ab}\wedge * (e_a\wedge e_b)
-2\lambda *1
+a_2\,T^{(2)}{}_a \wedge * T^a 
\right.
$$
$$
\left.
-2\left(c_3\,Q^{(3)}{}_{ab}+c_4\,Q^{(4)}{}_{ab}\right)
 \wedge e^a \wedge * T^b 
+\left(b_3\,Q^{(3)}{}_{ab}+b_4\,Q^{(4)}{}_{ab}\right) \wedge * Q^{ab}
\right]
$$
$$
-{z_4\over32}\, d Q \wedge * d Q
$$
where $\{\k,a_0,a_2,b_3,b_4,c_3,c_4,z_4\}$ are couplings and
$$
T^{(2)}{}^a\equiv {1\over3} e^a \wedge T
$$
$$
Q^{(3)}{}_{ab}\equiv
{4\over9}\left(
e_{(a}\,i_{b)}{\qq}-{1\over4}g_{ab}{\qq}
\right)
$$
$$
Q^{(4)}{}_{ab}\equiv{1\over4}g_{ab}Q
$$
$$
\qq\equiv e^a i^b \hat{Q}{}_{ab}
$$
$$
\hat{Q}{}_{ab}\equiv{Q}{}_{ab}-{1\over4}g_{ab}Q.
$$
We have verified that in accordance with our results the ansatz
\Qansatz\ and \Tansatz\ enables one to generate solutions to  this theory
 from an  Einstein-Proca type system with
unconstrained
cosmological constant $\lambda$ and arbitrary
couplings 
$\{\k,a_0$, $a_2$, $b_3$, $b_4$, $c_3$, $c_4$, $z_4\}$. Furthermore 
such an ansatz also  enables one to find solutions  by
 solving an Einstein-Maxwell 
type system (with or without $\lambda$) if these couplings 
satisfy the following relation:
$$
12\,a_0\,a_2\,b_3
+24\,{a_0}^2\,c_4
+6\,{a_0}^2\,a_2
+48\,a_0\,b_4\,c_3
+64\,a_0\,b_3\,b_4
-32\,a_2\,b_3\,b_4
+3\,a_0\,{c_4}^2
+18\,a_0\,c_3\,c_4
$$
$$
-9\,a_0\,{c_3}^2
-4\,a_0\,a_2\,b_4
+48\,a_0\,b_3\,c_4
+24\,b_4\,{c_3}^2
+24\,b_3\,{c_4}^2
+32\,{a_0}^2\,b_4
=0.
$$
\Section{Conclusions}

We have argued that with the aid of the ansatz \Qansatz\ and \Tansatz\
one may construct solutions with non-Riemannian fields  to
\LambdaCartaneqnsone,\ \LambdaCartaneqnstwo,\ \LambdaEineqns\
  from all solutions of Einstein-Maxwell type in
general relativity provided the couplings in the action described by
\LLambda\ are constrained according to \couplingcondition . 
This constraint can be relaxed and the action generalised to
\extendedaction\ provided solutions are then constructed
from any solution of Einstein-Proca type in
general relativity. 
In both cases one may generalise all field equations by adding an arbitrary
cosmological constant. 
These general results also apply to the actions discussed recently in  
\TresC\  and \TresD\  where particular solutions have been recognised.
We have also noted that there exists the more general ansatz
\starTansatz\ which generates  solutions containing
an arbitrary 1-form from this construction,
albeit with the further coupling restriction \couplingconditionA .

\Section{ Acknowledgment}

JS and RWT are  grateful to Middle East Technical University, Ankara 
for providing hospitality. 
TD, MO and RWT are grateful for the support of NATO grant CRG/941210
 and RWT 
to the Human Capital and Mobility Programme of the European Union for
partial support. CW is grateful to the School of Physics and Chemistry,
University of Lancaster  for a School Studentship, 
to the Committee of Vice-Chancellors and Principals, 
UK for an Overseas Research Studentship
and to the University of Lancaster for a Peel Studentship.

\vfill\eject

\References

\bye

\bye

\Section{ Acknowledgment}

RWT is grateful to R Kerner for providing facilities at the 
Laboratoire de Gravitation et Cosmologie Relativistes, Universite Pierre et
Marie Curie, CNRS,  Paris where this work was begun and
to the Human Capital and Mobility Programme of the European Union for
partial
support. CW is grateful
to the School of Physics and Chemistry,
University of Lancaster  for a School Studentship, 
to the Committee of Vice-Chancellors and Principals, 
UK for an Overseas Research
Studentship
and to the University of Lancaster for a Peel Studentship.
We are grateful to F Hehl for pointing out reference \added\ to us.

\vfill\eject

\References

\bye